\documentclass{ifacconf}

\usepackage{algorithm}

\usepackage{graphicx}      
\usepackage{natbib}        

\newtheorem{remark}{Remark}[section]

\begin{document}
\begin{frontmatter}

\title{Finite-Time Stabilization of Longitudinal Control for Autonomous Vehicles via a Model-Free Approach\thanksref{footnoteinfo}} 

\thanks[footnoteinfo]{This work was supported by the international Chair MINES ParisTech - Peugeot-Citro{\"{e}}n - Safran - Valeo on ground vehicle automation and the ANR project 15 CE23 0007 (Project Finite4SoS).}

\author[First]{Philip Polack}
\author[First]{Brigitte d'Andr\'ea-Novel}
\author[Second,Third]{Michel Fliess}
\author[First]{Arnaud de La Fortelle}  
\author[Fourth]{Lghani Menhour}

\address[First]{Centre de Robotique, Mines ParisTech, PSL Research University, \\ 60 boulevard Saint-Michel, 75006 Paris, France\\ (e-mail: \{philip.polack, brigitte.dandrea-novel, \\ arnaud.de\_la\_fortelle\}@mines-paristech.fr)}
\address[Second]{LIX (CNRS, UMR 7161), \'Ecole polytechnique, 91128 Palaiseau, France (e-mail: michel.fliess@polytechnique.edu)}
\address[Third]{AL.I.E.N., 24-30 rue Lionnois, BP 60120, 54003 Nancy, France (e-mail: michel.fliess@alien-sas.com)}
\address[Fourth]{Centre de Recherche STIC, Universit\'e de Reims Champagne-Ardenne, IUT de Troyes,\\  9 rue du Qu\'ebec, 10000 Troyes, France \\ (e-mail: lghani.menhour@univ-reims.fr)}

\begin{abstract}                
This communication presents a longitudinal model-free control approach for computing the wheel torque command to be applied on a vehicle. This setting enables us to overcome the problem of unknown vehicle parameters for generating a suitable control law. An important parameter in this control setting is made time-varying for ensuring finite-time stability. Several convincing computer simulations are displayed and discussed. Overshoots become therefore smaller. The driving comfort is increased and the robustness to time-delays is improved. 
\end{abstract}


\end{frontmatter}

\section{Introduction}

Research on autonomous vehicles has gained a growing interest since the successive DARPA challenges in 2004, 2005 and 2007. They are expected on the one hand to increase the road safety by taking the human out of the driving loop, and on the other hand to improve the efficiency of road transportation.

Among the robotic fields underlying in an autonomous vehicle, the low-level control is in charge of generating throttle/brake and steering commands of the vehicle. The input of a low-level controller can be either the output of a motion planner, meaning a list of waypoints $[x_{ref}, y_{ref}, v_{ref}]$ to follow, or the output of a higher-level controller, for example a speed reference and distance to the preceding vehicle. This work presents a longitudinal low-level controller where a reference target speed is given. It can be coupled with a regular motion planner such as a Rapidly-exploring Random Tree ($RRT$) (\cite{Kuwata2008}) or an $A^\star$ algorithm (\cite{Dolgov2008}) (see for example \cite{Katrakazas2015} for a full review on motion planning techniques), or a High-Level controller such as an Adapting Cruise Control (see \cite{Martinez2007}, \cite{Rajamani2012}) or a Stop-\&-Go system (\cite{Villagra2009}).

Model-based low-level controllers have been widely studied in the literature. A distinction is made between kinematic and dynamic controllers. In the first case, we can mention \cite{Samson1995}, \cite{Rajamani2012}, or the pure-pursuit controller introduced in \cite{Coulter1992}, and their variants. They are very performant but limited to low speed applications. They assume that the tire do not slip and skid. These assumptions are rather restrictive in the perspective of future fully automated autonomous vehicles running at high speed in an uncertain environment. 

Dynamic models are derived from the equations of the vehicle dynamics and those of the tires. These equations are highly nonlinear. Therefore, the generation of a control command is rather complicated. Linearisation assumptions, Linear Parameter Varying (LPV) system and flatness-based control techniques (\cite{Fliess1995}) have shown good results for regular driving situations but they suffer from a lack of robustness with regards to unknown dynamics such as the non-linear tire efforts or load transfers. In \cite{Menhour2013b}, their non-robustness to a change of parameter value such as mass or cornering stiffness is shown. Lastly, they are also sensitive to disturbances such as slopes, road-bank angle, aerodynamic forces or tire-road adhesion.

To bypass the nonlinearities of the dynamic equations, Model Predictive Control (MPC) is used in \cite{Camacho1999}, \cite{Falcone2007a}. The idea is to intertwine the planning and control parts by considering the dynamic equations as a constraint. The control input to apply is obtained through an optimisation planning problem. 

To tackle the problems of non-linearity and weak-robustness to parameter variation, low-level controllers based on model-free approach are popular. The well-known classic PID have been widely used (see \cite{Astrom1995}). Although this approach can be satisfying, no proof of stability can be obtained. A new model-free approach coupled with an ``intelligent" controller has been therefore introduced in \cite{Fliess2013}. It has been applied successfully to vehicle control in \cite{Menhour2013a}, \cite{Menhour2015} and \cite{DAndrea-Novel2016}. This method is very promising as it guarantees local asymptotic stability without any prior knowledge of our system. It captures the vehicle dynamics through online-estimation. The most rapid and violent movements which are considered here lead us to employ a recent important advance on a time-varying approach due to \cite{Doublet}. This improvement leads moreover to finite-time stabilization and allows to tackle severe speed discontinuities little covered in existing literature (\textit{e.g.} step function). 

This communication is organized as following. Section \ref{section:MFC} presents the model-free approach defined by \cite{Fliess2013}. Section \ref{section:stability}  exploits the approach by \cite{Doublet} in order to obtain a finite-time stability. The 7 Degrees of Freedom (DoF) simulation model used to test our controller is presented in Section \ref{section:simu_model}. Section \ref{section:results} presents the results obtained and compares them to the classic model-free control approach without adaptation. Some concluding remarks and future investigations are presented in Section \ref{section:conclusion}.

\section{Longitudinal Model-Free Control}
\label{section:MFC}
\subsection{Short review on model-free control}

The model-free control approach presented here was introduced by \cite{Fliess2013}. It has been applied to the automation of vehicle in several papers such as \cite{Menhour2013a}, \cite{Menhour2015} and \cite{DAndrea-Novel2016}.\\

The longitudinal dynamics of the vehicle is approximated by an ultra-local differential
relation of order 1 valid only on a very short time period ($T \approx 200$ ms):
\begin{eqnarray}
	\label{eq:ultralocal_MFC}
	\dot{y} & = & F + \alpha u 
\end{eqnarray}
where $y$ is the control output, $F$ represents both the unmodelled and the neglected dynamics, $u$ is the control input, and $\alpha$ is a parameter chosen by the practitioner such that $\alpha u$ has the same order of magnitude as $F$.
The model-free approach has two main assets:
\begin{itemize}
	\item No prior knowledge of vehicle parameters and tire characteristics is needed (for ex. $l_f$, $l_r$, $M_T$ - see notation in Table~\ref{tb:notations}),
	\item The whole dynamics of the car (even the unmodelled one) and the external disturbances such as wind, slope, road-bank are taken into account.
\end{itemize}

In the case of a longitudinal control, we choose the longitudinal speed $V_x$ as control output $y$ and the sum $C_T$ of torque applied at each wheel as control input $u$, such as in \cite{Menhour2015}. A system of first order was chosen, based on the dynamic analytical model of a vehicle:

\begin{eqnarray}
	\dot{V}_x & = & F + \alpha C_T
\end{eqnarray} 

\begin{remark}
	The parameter $\alpha$ is nonnegative as the longitudinal speed of a vehicle is an increasing function of the applied torque on the wheels.
\end{remark}

\subsection{Estimation of the unmodelled and neglected dynamics}

The unmodeled and neglected dynamics at time t, $F(t)$, can be estimated using directly the last applied control  input $u(t-1)$ and the last observed output $y(t)$:

\begin{eqnarray}
	\widehat{F}(t)= \dot{y}(t)-\alpha u(t-1)
\end{eqnarray}

However, the raw signals such as speeds are often rather noisy. A  filter has been derived from \cite{Fliess2003}. Based on operational calculus (see, \textit{e.g.}, \cite{yosida}), it is used in order to estimate $\widehat{F}$:

First, we transform Equation~(\ref{eq:ultralocal_MFC}) in the operational domain:

\begin{eqnarray}
	sY(s) & = & \frac{F_s}{s}+\alpha U(s)+ y(0)
\end{eqnarray}
where $F_s$ is constant. Getting rid of the initial condition by multiplying by $\frac{d}{ds}$ on both sides, multiplying then by $s^{-2}$ in order to achieve noise attenuation yields:

\begin{eqnarray}
	\frac{Y(s)}{s^2} + \frac{1}{s}\frac{dY}{ds} & = - & \frac{F_s}{s^{-4}}+\alpha \frac{1}{s^2}\frac{dU}{ds}
\end{eqnarray}
From the above algebraic calculations we get 
\begin{eqnarray}\label{Eqn:filter}
	\widehat{F} & = & - \frac{6}{T^3} \int_{0}^{T} [(T-2\tau)y(\tau)+\alpha\tau(T-\tau)u(\tau))]\, \mathrm{d}\tau
\end{eqnarray}	
There we exploit the connection between $\frac{d}{ds}$ and the multiplication by $- t$ in the time domain. Going from Equation (\ref{Eqn:filter}) to a digital filter is done using the last $(\lfloor T/\Delta t\rfloor+1)$ values of $y$ and $u$, where $\Delta t$ is the sampling time of the controller, and $T$ the filtering window.

\subsection{Intelligent Controller}

Lastly, the next control input $u(t)$ to apply at time t is computed from the estimation $\widehat{F}(t)$, the reference longitudinal speed $y_{r}$ and the error on longitudinal speed $e(t)=y(t)-y_r(t)$:

\begin{eqnarray}
	\label{eq:control_MFC}
	u & = & -\frac{\widehat{F}-\dot{y}_r+K_Pe}{\alpha}	
\end{eqnarray}
It yields
\begin{eqnarray}
	\label{eq:pde_error}
	\dot{e}(t)+K_p e(t) & = & e_F = F - \widehat{F}
\end{eqnarray}
Choosing $K_P > 0$ is fine if $e_F$ is ``small'', \textit{i.e.}, if the estimate $\widehat{F}$ is ``good''.


\section{Adaptive model-free control}
\label{section:stability}
The fast and violent movements which are considered here lead us to introduce an \emph{adaptive model-free} control setting, where, according to \cite{Doublet}, the coefficient $\alpha$ in Equation (\ref{eq:ultralocal_MFC}) becomes a piecewise continuous function $\alpha(t)$, updated at each time step. 

\begin{remark}
The wording ``adaptive model-free control'' is encountered elsewhere, but with a different meaning (see, \textit{e.g.}, \cite{battistelli,hou,roman}, and the references therein). 
\end{remark}








From Equation~(\ref{eq:control_MFC}), the error $e(t)$ may be expressed in the following way:
\begin{eqnarray}
	e(t) & = & \frac{1}{K_p}\left(-\widehat{F}+\dot{y}_r(t)-{\alpha(t)}u(t) \right) 
\end{eqnarray}

Choose $\alpha(t)$ such that $e(t)=0$ in order to achieve finite-time stability: 

\begin{eqnarray}
	\alpha(t) & = & \frac{-\widehat{F}+\dot{y}_r(t)}{u(t)}
\end{eqnarray}


In order to avoid singularities and nonpositive values of $\alpha(t)$, set 

\begin{eqnarray}
\label{eq:alpha_impl}
	\widehat{\alpha}(t) & = & \max \left(\frac{-\widehat{F}+\dot{y}_r(t)}{u(t)+\epsilon sign(u(t))}, \alpha_{\rm{nominal}} \right)
\end{eqnarray}
where $\alpha_{\rm{nominal}} > 0$ is selected by the practitioner manually via trial and error, $sign$ is the sign function (with $sign(0)=+1$) and $\epsilon=0.01$.

\begin{remark}
The filter in Equation (\ref{Eqn:filter}) remains valid on $[kT;(k+1)T]$ if we consider the following system: ${\dot{y}(t)=F(t)+V(t)}$ where $V(t)=\alpha(t)u(t)$. This leads to the constant $\alpha$ case of Equation~(\ref{eq:ultralocal_MFC}), with $\alpha=1$. More precisely, Algorithm (\ref{alg:1}) is used.	
\end{remark}

\begin{alg}
\label{alg:1}
At each time step $t_k$:\\
$\widehat{F}_k =$filter$(\alpha=1,v_{k-1}=\widehat{\alpha}_{k-1}u_{k-1}, y_k)$ given by (\ref{Eqn:filter})\\ 
$u_k=$iP$(\widehat{F}_k,\widehat{\alpha}_{k-1}, y^{ref}_k)$ given by (\ref{eq:control_MFC})\\
$\widehat{\alpha}_k=\max \left(\frac{-\widehat{F}_k+ \dot{y}^{ref}_k}{u_k+\epsilon sign(u_k)}, \alpha_{\rm{nominal}}\right)$ given by (\ref{eq:alpha_impl})\\ 
\end{alg}

\begin{remark}
Let us point out that Equation (\ref{eq:pde_error}) remains valid with $\alpha$ varying in Equation (\ref{eq:control_MFC}).
\end{remark}

\section{Model for Simulations}
\label{section:simu_model}

In our simulations, several assumptions are made:
\begin{itemize}
	\item We neglect the external disturbances such as aerodynamic forces, road-bank angle and slope.
	\item The dynamics of the car engine is not taken into account.
\end{itemize}

\begin{figure}[h]
\begin{center}
\includegraphics[scale=0.30]{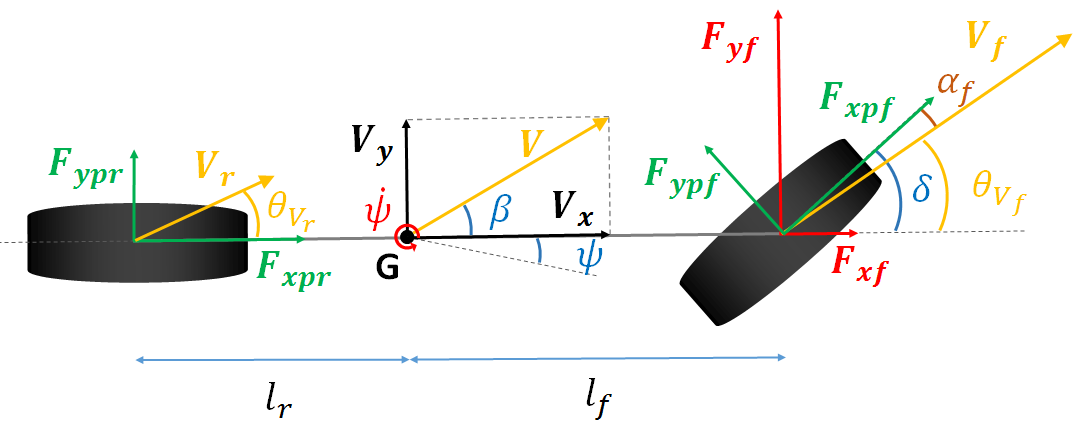}
\caption{7 DoF vehicle model used for simulations} 
\label{fig:7DoF}
\end{center}
\end{figure}

The notations used are summarized in Table~\ref{tb:notations} and Figure~\ref{fig:7DoF}.

\begin{table}[hb!]
\begin{center}
\caption{Notations}\label{tb:notations}
\begin{tabular}{l|p{6.6cm}}
\hline
Symbols & Variables\\\hline
\hline
$M_T$ & vehicle mass [kg]\\
$l_f$, $l_r$ & distances from the center of gravity to the front and rear axles [m]\\
$I_z$ & vertical inertia of the vehicle [kg.m$^{-2}$]\\
$I_r$ & wheel inertia [kg.m$^{-2}$]\\
$r_{eff}$ & effective radius of the wheel [m]\\
$V_x$, $V_y$ & longitudinal and lateral speed [m/s]\\
$\dot{\psi}$, $\psi$ & yaw rate [rad/s] and yaw angle [rad]\\
$u=C_T$ & wheel torque [Nm]\\
$C_{mi}$, $C_{fi}$ & motor and brake torque applied at wheel $i$ [Nm]\\
$F_{xf}$ , $F_{xr}$ & longitudinal forces in the vehicle coordinates of front and rear wheels [N]\\
$F_{yf}$ , $F_{yr}$ & lateral forces in the vehicle coordinates of front and rear wheels [N]\\
$F_{xpi}$ & longitudinal forces applied on wheel $i$ in the tire coordinates [N]\\
$\tau_x$ & longitudinal slip ratio\\
\hline
\end{tabular}
\end{center}
\end{table}

\subsection{7 DoF vehicle model}

A 7 Degrees of Freedom (DoF) vehicle model is used for designing and testing purposes. This is a suitable model of a car moving on a plane (vertical effects are not taken into account). The forces that the road applies on each tire is modelled through Pacejka's ``Magic Tire Formula" (see \cite{Pacejka1997}).

\subsubsection{Vehicle dynamics}
The 7 DoF of the vehicle model used for simulation are: longitudinal speed $V_x$, lateral speed $V_y$, yaw speed $\dot{\psi}$ and the rotating speed of each wheel  $\omega_i$.

The equations of the car-body moving on a plane can be written (see table~\ref{tb:notations} for notation):
\begin{eqnarray}
	\label{Eqn_3DDL-1}
	M_T(\dot{V}_x-\dot{\psi}V_y) & = & F_{xf}+F_{xr}\\ 
	\label{Eqn_3DDL-2}	
	M_T(\dot{V}_y+\dot{\psi}V_x) & = & F_{yf}+F_{yr}\\
	\label{Eqn_3DDL-3}
	I_z \ddot{\psi} & = & l_f F_{yf}-l_r F_{yr}
\end{eqnarray}

The dynamic of each wheel $i$ can be written:
\begin{eqnarray}
	\label{Eqn_dyn_roues}
	\dot{\omega}_i=\frac{C_{mi}-C_{fi}-F_{xpi}\times r_{eff}}{I_{ri}}
\end{eqnarray}

\subsubsection{Pacejka tire dynamics}

The tire model used in our simulation is given by Pacejka's formula in \cite{Pacejka1997}. In its easiest form, the longitudinal forces $F_{xp}$ depend only on the tire longitudinal slip ratio $\tau_x$:
\footnotesize
\begin{eqnarray}
	\label{Eqn:Pacejka}
	F_{xp}(\tau_x) & = D \sin[C\arctan(B\tau_x-E(B\tau_x-\arctan(B\tau_x)))]
\end{eqnarray}
\normalsize


The expression of the longitudinal slip ratio depends on the velocities of the vehicle and of the tire:
\begin{itemize}
	\item during propulsion: $\tau_x = \frac{r_{eff} \omega - V_x}{r_{eff} \omega}$\\
	\item during braking: $\tau_x = \frac{r_{eff} \omega - V_x}{|V_x|}$\\
\end{itemize}
In the case where the longitudinal speed $V_x$ and the rotation of the tire $\omega$ are in the opposite direction, the value of $\tau_x$ is saturated to $[-1;1]$.

$B$ is called the stiffness factor, $C$ the shape factor, $D$ the peak value of $F_{xp}$ and $E$ the curvature factor. These constants can be computed from the experimental curve $F_{xp}(\tau_x)$ in the following manner:
\begin{enumerate}
	\item $D$ is the peak value of $F_{xp}$;
	\item $C=2-\frac{2}{\pi}\arcsin(\frac{y_s}{D})$ where $y_s$ is the asymptotic value of $F_{xp}$;
	\item $B$ is computed from the initial slope egal to $BCD$;
	\item $E=\frac{B \tau_{x,m} - \tan(\frac{\pi}{2C})}{B \tau_{x,m} - \arctan(B \tau_{x,m})}$ where $\tau_{x,m}$ is the value such that $F_{xp}(\tau_{x,m})=D$.
\end{enumerate}

In reality, they are not real constants but depend on the friction coefficient $\mu$ and the normal reaction $F_z$ of the road on the tire.

A similar model is used for lateral forces $F_{yp}$ in the tire coordinates with regards to slip angle.

\subsubsection{Adding Noise}

Our simulator gives us the ``ground truth" about the states of the vehicle. Therefore, in order to make our simulations more realistic and test the robustness of our control law, we added a -6dB white Gaussian noise to the simulated longitudinal speed $V_x$. Thus, the speed signal is closer to the one measured or estimated using sensors on real vehicles. 


\section{Results}
\label{section:results}

In order to test our newly introduced adaptive longitudinal controller, three tests were performed with different speed reference inputs: a step function reference speed, a sinusoidal function reference speed and a real driver speed input performed on a track. To the best of our knowledge, such challenging tests with severe speed discontinuities have not been frequently studied in the literature. A comparison is made between the classic model-free control (Classic MFC) presented in \cite{Menhour2015} and the proposed adaptive model-free control approach (Adaptive MFC), using for $\alpha_{\rm{nominal}}$ the constant value $\alpha$ of the classic approach.

\subsection{Response to a step function}

In this case, the car is moving on a straight line and the reference speed is a step function of position. This is a challenging speed to follow for a longitudinal controller due to the non regularity of the speed reference.

\begin{figure}[h!]
\begin{center}
\includegraphics[]{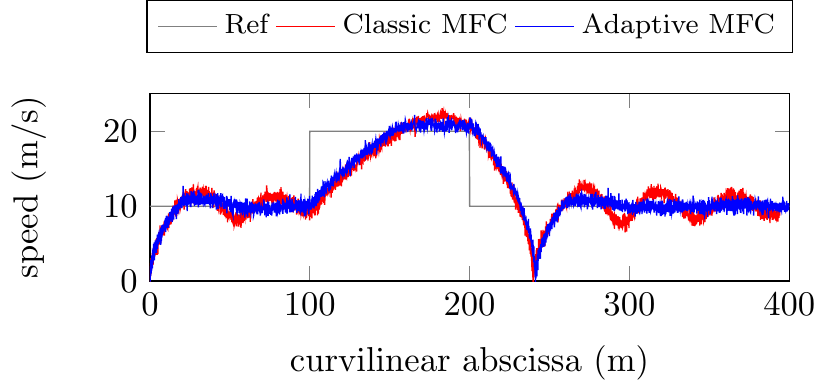}
\caption{Comparison of the longitudinal speeds between classic and adaptive model-free control to successive steps in target speed} 
\label{fig:S1_speed}
\end{center}
\end{figure}

\begin{figure}[h!]
\begin{center}
\includegraphics[]{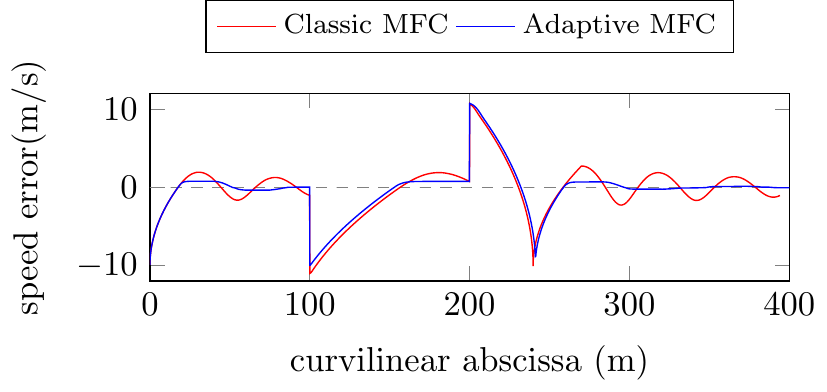}
\caption{Comparison of the longitudinal speed errors (ground truth) between classic and adaptive model-free control to successive steps in target speed} 
\label{fig:S1_error}
\end{center}
\end{figure}

\begin{figure}[h!]
\begin{center}
\includegraphics[]{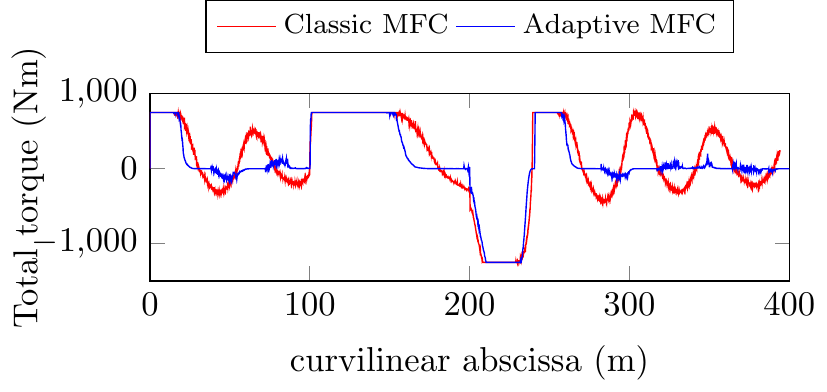}
\caption{Comparison of the computed motor/brake torque control input for one of the front wheel between the classic and adaptive model-free approach to successive steps in target speed} 
\label{fig:S1_control}
\end{center}
\end{figure}

The results obtained with a tuned classic model-free control approach and an adaptive model-free control approach are shown on Fig.~\ref{fig:S1_speed} and Fig.~\ref{fig:S1_error}. In the classic approach, we observe that the system is asymptotically stable but oscillatory and the overshoot is rather important: for the first step, we have about 19.5\% overshoot and for the second around 9.5\%. In comparison, our adaptive model-free control approach shows better results as the system is less oscillatory and the overshoot is greatly reduced: for the first step in target speed, the overshoot is around 8\% and for the second, around 3.9\%. Therefore, our adaptive model-free control approach is more robust to jumps in target speed compared to existing model-free approaches. Moreover, the system stabilises itself in finite-time, between 50m and 100m (see Fig.~\ref{fig:S1_error}), ensuring a smoother driving. This is confirmed by the control input (see Fig.~\ref{fig:S1_control}) that stabilizes itself quickly around the value $0$. 

\begin{figure}[h!]
\begin{center}
\includegraphics[]{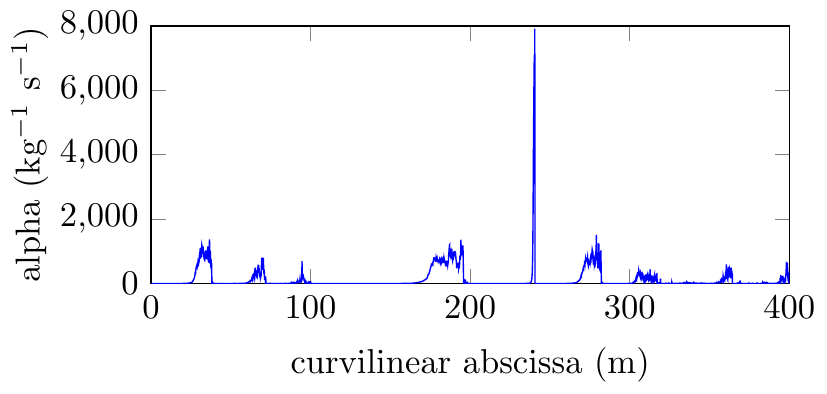}
\caption{Evolution of the estimation of $\widehat{\alpha}(t)$ with the curvilinear abscissa for successive steps in target speed} 
\label{fig:S1_alpha}
\end{center}
\end{figure}

Fig.~\ref{fig:S1_alpha} shows the evolution of $\widehat{\alpha}(s)$ with respect to the curvilinear abscissa. We observe that its value increases strongly when the speed of the vehicle exceeds the target one: this implies a reduction of the torque input, thus reducing the overshooting and avoiding oscillations. Please note that the value of $\widehat{\alpha}(s)$ is $\alpha_{\rm{nominal}}=1$ most of the time, and not $0$.

\subsection{Response to a sinusoidal function}

In this case, the car is moving on a straight line and the reference speed is a sinusoidal function of position. This time, the target speed is smooth but keeps on varying. Therefore, the fast convergence of the longitudinal error is important.

\begin{figure}[h!]
\begin{center}
\includegraphics[]{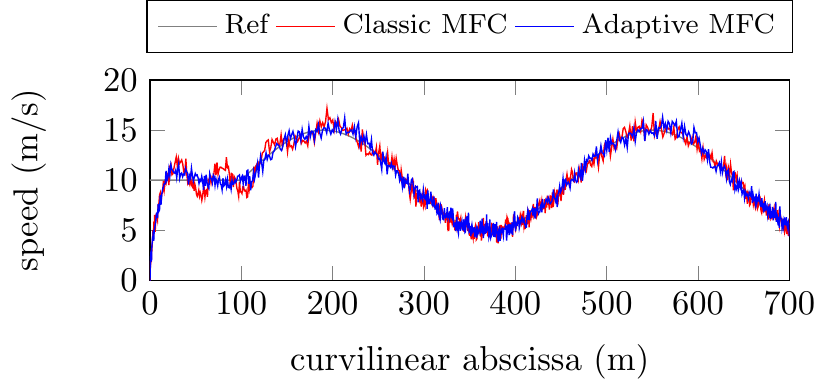}
\caption{Comparison of the longitudinal speeds between classic and adaptive model-free control to a sinusoidal target speed} 
\label{fig:S2_speed}
\end{center}
\end{figure}

\begin{figure}[h!]
\begin{center}
\includegraphics[]{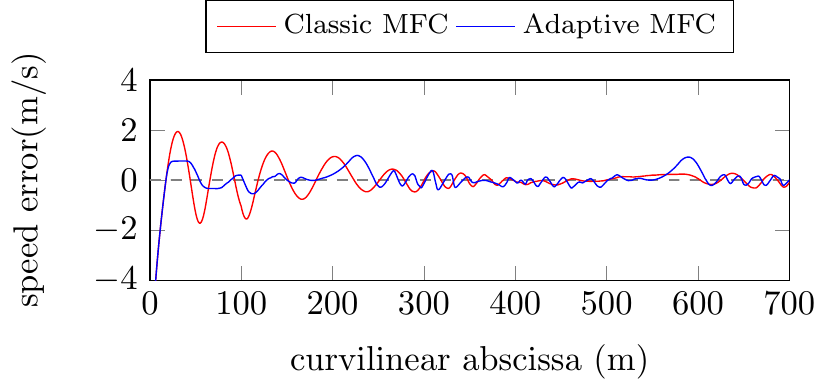}
\caption{Comparison of the longitudinal speed errors (ground truth) between classic and adaptive model-free control to a sinusoidal target speed} 
\label{fig:S2_error}
\end{center}
\end{figure}

We observe on Fig.~\ref{fig:S2_speed} and Fig.~\ref{fig:S2_error} that the classic model-free control and our adaptive longitudinal control give approximately the same results. However, the convergence of our system is quicker on the first oscillation and the overshooting is almost non-existing.

\subsection{Response to reference speed inputs acquired from real driver data}

Lastly, we used as reference speed input some driver data recorded on a real track to compare the two model-free approaches. The recorded data were the state of the vehicle $[V_x,V_y, \dot{\psi}]$ and the time $t$. The positions of the waypoints $[x_{ref},y_{ref}]$ were thus obtained by reconstruction using the following equations:

\begin{eqnarray}
	\Delta_t & \leftarrow & t(i+1)-t(i)\\
	\Delta_x & \leftarrow & x_{ref}(i+1)-x_{ref}(i)\\
	\Delta_y & \leftarrow & y_{ref}(i+1)-y_{ref}(i)\\
	x_{ref} & \leftarrow & x_{ref} + V_x\Delta_t \cos(\psi_{ref})-V_y\Delta_t \sin(\psi_{ref})\\
    y_{ref}& \leftarrow & y_{ref} + V_x\Delta_t \sin(\psi_{ref})+V_y\Delta_t \cos(\psi_{ref})\\
	\psi_{ref} & \leftarrow &  \psi_{ref}+ \Delta_t \dot{\psi}_{ref};\\
    s_{ref} & \leftarrow & s_{ref}+\sqrt{\Delta_x^2+\Delta_y^2};
\end{eqnarray}

where $V_x$ and $V_y$ were filtered from the raw data (average over 100 points) and $s_{ref}$ is the curvilinear abscissa reference of the waypoints.

\begin{figure}[h!]
\begin{center}
\includegraphics[]{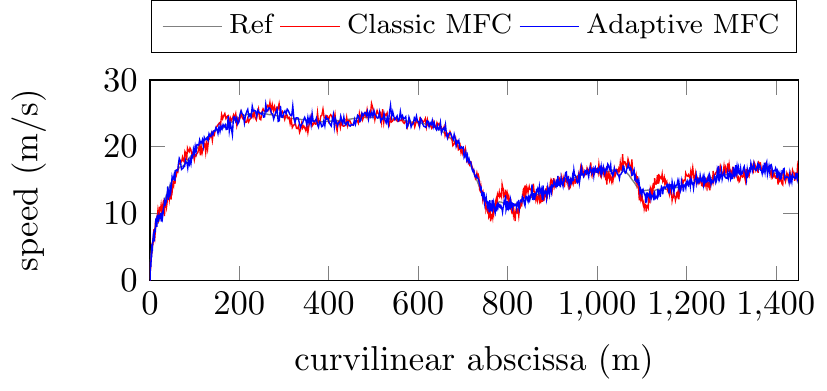}
\caption{Comparison of the longitudinal speeds between classic and adaptive model-free control on real driver reference data}
\label{fig:S3_speed}
\end{center}
\end{figure}

\begin{figure}[h!]
\begin{center}
\includegraphics[]{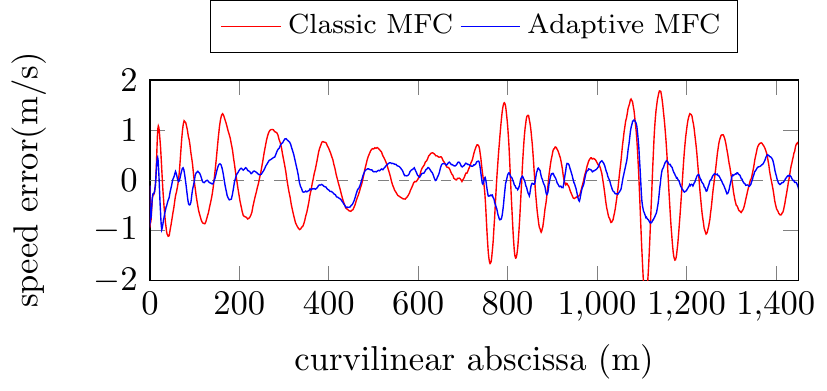}
\caption{Comparison of the longitudinal speed errors (ground truth) between classic and adaptive model-free control on real driver reference data} 
\label{fig:S3_error}
\end{center}
\end{figure}

Comparing both models in Fig.~\ref{fig:S3_speed} and Fig.~\ref{fig:S3_error}, we observe that the adaptive model-free approach is more stable: the results are less oscillatory and give smaller longitudinal errors. Table~\ref{tb:S3_data_error} shows in fact that the controller is able to follow the reference target speed with a better precision (smaller root mean square error) and with less oscillations (smaller standard deviation).

\begin{table}[hb!]
\begin{center}
\caption{Comparison between the error of classic MFC and adaptive MFC on real driver reference data}\label{tb:S3_data_error}
\begin{tabular}{c|c|c|c}
\hline
MFC & Average & Std. dev. & Root mean square\\
\hline
\hline
adaptive & -0.02 & 0.35 & 0.35\\
classic & -0.01 & 0.77 & 0.78\\
\hline
\end{tabular}
\end{center}
\end{table}

\subsection{Robustness to delays}

The results presented in \cite{Doublet} suggest that the adaptive model-free approach presented in this paper is more robust to delays. This is an important characteristic of our system as delays could arise from the filter presented in Equation (\ref{Eqn:filter}) and the car engine dynamics. Therefore, we ran simulations with a $250$ms time delay on the control input for both the classic and adaptive MFC on the real driver data. The results showed that adapting $\alpha$ gives better results: the root mean square of the error was only $0.68$m/s for the adaptive MFC (quite similar to the classic MFC without delay) and $2.27$m/s for the classic MFC.


\section{Conclusion}
\label{section:conclusion}

We have introduced an adaptive longitudinal model-free control. The online-line adaptation of the parameter $\alpha$ enables to stabilize the system in finite time and to smooth the control input. Adaptation happens when the sign of the longitudinal error changes. This leads to reduce the value of the control input as well as the time response of the closed-loop system. Therefore the comfort of the driver is improved, especially when the reference speed changes abruptly. It also drastically reduces the control input overshoot which is important for respecting speed limit and maintain safety. Lastly, the adaptive model-free control is more robust to time delays arising in the execution of the control input. These delays are very likely to exist on real vehicles due to the car engine dynamics. In a future work, we will extend the approach to the lateral control of the vehicle.

\begin{ack}
The authors would like to thank warmly C\'edric Join and Maxime Doublet for useful discussions concerning the paper \cite{Doublet}, and the adaptation of the control parameter.
\end{ack}

\bibliography{ifacconf}             

\end{document}